\documentclass[11pt]{article}
\usepackage{moriond,epsfig, amssymb}
\usepackage{axodraw,psfrag}

\def\be{\begin{equation}}
\def\ee{\end{equation}}
\def\bea{\begin{eqnarray}}
\def\eea{\end{eqnarray}}

\begin{document}
\vspace*{4cm}
\title{Electroweak Symmetry Breaking in Warped Extra Dimensions}

\author{ Gero von Gersdorff }

\address{CPhT, Ecole Polytechnique,\\
91128 Palaiseau CEDEX, France}

\maketitle\abstracts{
We review Electroweak breaking in warped Extra Dimensions and show how it is constrained by Electroweak precision tests. We then proceed to describe a model which is based on a bulk Higgs field and a metric that is deformed from the usual five dimensional Anti de Sitter space near the Infrared (IR) boundary. It allows to softly decouple the Higgs field from the Kaluza Klein excitations near the IR and thus reduces their contribution to the precision observables.}

\section{The RS model and its variants}

Following the influential work of Randall and Sundrum,\cite{Randall:1999ee} in the last decade a lot of effort has been put into constructing realistic warped models of Electroweak Symmetry breaking (EWSB) which obey all experimental constraints obtained in precision measurements of electroweak and flavour observables.
In the Randall-Sundrum (RS) setup, a five dimensional (5D) bulk is endowed with an Anti de Sitter (AdS) metric that in proper coordinates is given by
\be
ds^2=e^{-2ky}\,\eta_{\mu\nu}\,dx^\mu dx^\nu + dy^2\,,
\label{AdS}
\ee
with the Minkowski metric $\eta_{\mu\nu}$ with signature ($-$$+$$+$$+$). The constant $k$ of dimension mass is the inverse AdS curvature radius and is considered to be of the order of the Planck scale.  
The presence of the warp factor $e^{-ky}$ introduces a scale dependence along the extra dimension and causes all mass scales to be redshifted when moving along increasing $y$. 
The space is bounded by two "branes", located at $y=0$ (UV brane) and $y=y_1$ (IR brane).

In the simplest setup, the Higgs field is localized at the IR boundary:
\be
\mathcal L_{\rm brane-Higgs} =\sqrt{g}\,  \left(-g^{\mu\nu}\, D_\mu H^\dagger D_\nu H
-m_h^2 |H|^2
\right)\big|_{y=y_1}
=-e^{-2ky_1}\,  |D_\mu H|^2
-e^{-4ky_1}\, m_h^2 |H|^2\,.
\ee
Canonically normalizing this Lagrangian,  the -- originally Planck size -- Higgs mass $m_h$ is "warped down" to the Electroweak (EW) scale 
\be
m_h\to e^{-ky_1}\,m_h
\ee
provided the volume (in units of $k$) is about $ky_1=\log 10^{16}\sim 35$. This is a moderately large number that can be easily achieved dynamically by an appropriate stabilization mechanism.~\cite{Goldberger:1999uk}
The stabilization mechanism is also needed in order to give mass to the radion, the particle related to the fluctuations of the interbrane distance. Typically, the mass of the radion turns out to be about an order of magnitude lighter than the first KK excitations, although heavier masses are possible.~\cite{Cabrer:2011fb}

The smoking gun signature of the RS model is the presence of strongly coupled KK resonances of the graviton. By considering fluctuations around the metric Eq.~(\ref{AdS}) it is found that the graviton KK spectrum is quantized with TeV spacing, the lowest excitations being
\be
m^{\rm grav}_{\rm KK}=3.8\, k\, e^{-ky_1}\,.
\ee
Moreover, its coupling is not Planck suppressed but rather set by the IR scale. It is thus a universal prediction of all variants of the RS model to produce spin--2 resonances at LHC or other future particle colliders.
The wave functions of the KK modes are localized towards the IR boundary.

There exists a plethora of variations of this simplest setup, where the SM gauge and matter fields as well as the Higgs field are propagating in the bulk.~\cite{Gherghetta:2000qt} In this case, zero modes for the matter and Higgs fields typically feature nontrivial wave functions that depend on the bulk mass parameter. For instance, the Higgs boson zero mode has a profile
\be
h(y)\sim e^{ak(y-y_1)}\,, 
\ee
where the real parameter $a$ is related to the bulk Higgs mass term $M_h^2\,|H^2|$ as $M_h^2=k^2 a(a-4)$.
Inserting this profile back into the 5D Higgs Lagrangian, 
\be
\mathcal L_{\rm bulk-Higgs}=\int_0^{y_1}\,dy\,\sqrt{g}\left(-
g^{MN}\, D_M H^\dagger D_N H
- \left[
M_h^2+M_0\,\delta(y)-M_1\,\delta(y-y_1)\right]\,|H|^2\,
\right)\,,
\ee
one finds the physical Higgs mass %
\be
m_h^2=\frac{2(a-1)}{e^{2(a-1)ky_1}-1}\ \left((M_0-ak)k-(M_1-ak)ke^{2(a-2)ky_1}\right)\,.
\label{mh}
\ee
For $a<1$ this mass is $\mathcal O(k)$ unless one fine tunes $M_0=ka$. This is the original hierarchy problem. In fact, in this range the Higgs can be considered near-UV localized in the sense that its kinetic term is multiplied by $e^{2(a-1)ky}$.
For $1<a<2$, the hierarchy problem still partially persists, although the "natural " Higgs mass is now of the order $m_h\sim k\,e^{-(a-1)ky_1}$. Finally, for $a>2$, the Higgs mass is of the order $m_h\sim ke^{-k y_1}$ and the hierarchy problem is fully solved. For the pure AdS background metric, the Higgs mass is thus constrained by $a>2$ in order to fully address the hierarchy problem. There is a simple alternative interpretation of this result by invoking the AdS-CFT correspondence: The parameter $a$ is related to the dimension of the strongly coupled Higgs condensate via $\dim \mathcal O_H=a$. The condition $a>2$ thus renders the mass operator $\mathcal |O_H|^2$ irrelevant, which is precisely what is needed in order to solve the hierarchy problem.  It has also been shown that under certain conditions, gauge coupling unification can be achieved at a scale $M_{\rm GUT}\sim 10^{15}$ GeV with a precision comparable to that of the MSSM.~\cite{Agashe:2005vg}

We should mention two more possibilities for the Higgs sector, which have been employed in the context of warped extra dimensions. The first one goes by the name of gauge-Higgs unification, or under its dual alias "composite pseudo Nambu-Goldstone Higgs".~\cite{Agashe:2004rs} Instead of a fundamental Higgs field, one introduces an extended gauge group $G$ in the bulk, which by the IR boundary conditions is broken down to a subgroup $H$ containing the Standard Model gauge group. At the UV brane, only the SM survives the boundary conditions.  By appropriately choosing~\footnote{In the minimal model~\cite{Agashe:2004rs} $G=SO(5)\times U(1)$, $H=SU(2)_L\times SU(2)_R\times U(1)\supset SU(2)_L\times U(1)_{Y}$.} $G$ and $H$, the coset $ G/H$ contains scalars with the quantum numbers of the Higgs boson in the $A_5$ sector of the theory (the fifth component of the gauge boson). At tree level, its potential is flat due to the underlying 5D local gauge symmetry. However, at one loop a potential is generated~\cite{hosotani} due to a Wilson line needed to make all bulk propagators gauge covariant. The potential is nonlocal and finite. These models have the  advantage that the Higgs mass can be naturally light due to the vanishing tree level contribution. In the holographic dual version of the theory the Higgs is identified as the (pseudo-) Nambu Goldstone bosons of the breaking of the {\em global} symmetry $G\to H$. Recall that according to the AdS-CFT dictionary, 5D gauge symmetries correspond to global currents in the 4D dual. The fate of the current depends on the imposed boundary condition according to table \ref{tab}.
We see that the global symmetry $G$ is spontaneously broken to $H$ producing Goldstone bosons in the coset $G/H$. The SM, a subgroup of $H$, is gauged and further spontaneously broken by a VEV of the pseudo-Goldstone Higgs. The second alternative, the "Higgsless model",~\cite{Csaki:2003dt} is an extreme version of the previous one, in which the coset $G/H$ only contains the Goldstone bosons and no radial, "Higgslike" excitations. This can be achieved, for instance, by choosing $G=SU(2)_L\times SU(2)_R$ and $H=SU(2)_V\supset U(1)_{EM}$. Holographically these models thus bear some similarities to large-$N_c$ Technicolor models.
 
\begin{table}[t]
\caption{The properties of the global currents, depending on the boundary conditions imposed on them at the two branes.\label{tab}}
\vspace{0.4cm}
\begin{center}
\begin{tabular}{|c|cc|}
\hline
& &  \\
&Neumann&Dirichlet\\
\hline
UV brane &gauged&ungauged\\
IR brane &exact&spontaneously broken\\
&&\\
\hline
\end{tabular}
\end{center}
\end{table} 
 
\begin{figure}[t]
\begin{center}
\includegraphics[width=0.50\textwidth]{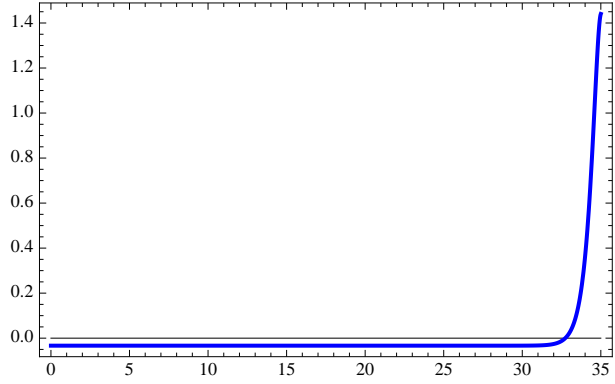}
\end{center}
\caption{The normalized wave function of the first gauge boson KK mode as a function of the coordinate $y$.}
\label{KK}
\end{figure}

Besides the aforementioned KK graviton modes which are universally present in any variant of the RS model, there are more model dependent resonances such as KK modes of the gauge bosons or even of the matter fields. These resonances can contribute to higher dimensional operators when integrated out at tree level and beyond, and hence precision measurements from LEP and flavour experiments tightly constrain the parameters of the model. Before turning to details of these constraints, let us pause and figure out the parameteric dependence of the coupling of SM zero modes to heavy KK resonances. The wave functions of the latter can be obtained by solving the 5D equations of motion (EOM) in the AdS background. It turns out that the wave functions of the first heavy KK resonances are mostly constant throughout the bulk and concentrated near the IR brane, see Fig.~\ref{KK}. The normalized wave function parametrically behaves as
\be
f_n(0)\sim (ky_1)^{-1}\,,\qquad f_n(y_1)\sim 1\,.
\label{vertex}
\ee
The SM fields couple to the KK resonances of the gauge fields as
\be
\mathcal L_{\rm SM-KK}=g_n\, J^\mu_{\rm SM}(x) \cdot A_\mu^n(x)\,,
\ee
where $J_{\rm SM}^\mu$ are the usual SM matter and Higgs currents. However, due to the non-flatness of the gauge-KK wave functions, the couplings can be highly nonuniversal, depending on the profile of the matter zero modes. Schematically,
\be
g_n=g_{5}\,\int f_n(y)\psi(y)^2 \sim \left\{\begin{array}{ll}
g_4\,(ky_1)^{-\frac{1}{2}}&{\rm for\ mostly\ UV\ localized\ fields,}\\
g_4\,(ky_1)^{+\frac{1}{2}}&{\rm for\ mostly\ IR\ localized\ fields,}
\end{array}
\right.\label{gn}
\ee 
where $\psi(y)$ are the zero mode wave functions normalized as $\int\, \psi(y)^2=1$,  and the 5D and 4D gauge couplings are related as $g_5=g_4 \sqrt{y_1}$.

\section{Electroweak precision tests for RS}

Any theory aiming to explain EWSB has to confront the precise measurements from LEP. In general, new physics beyond the SM will generate higher dimensional operators which are suppressed with inverse powers of the masses of the new states. In a completely model independent approach, one would like to classify these new operators, compute their contribution to the LEP observables and perform a global fit to their coefficients. 
It turns out that in a large class of models, including most versions of RS, there is only a certain subset of operators which are relevant. These are the so-called oblique corrections, and they are defined as follows
\be
\mathcal L_{\rm oblique}= \frac{1}{2\,m_W^{2}}\left(g^2\,\hat T\ |H^\dagger D_\mu H|^2
+gg'\,\hat S\ [H^\dagger\,W_{\mu\nu}H B^{\mu\nu}]+W\ [D_\rho W_{\mu\nu}]^2+Y\ [\partial_\rho B_{\mu\nu}]^2
\right)\,.
\label{oblique}
\ee
In the following, we will restrict ourselves to this subset.~\footnote{Strictly speaking for RS models one should add one more operator related to the modified $Zb\bar b$ coupling. We will comment on this operator below.} Note that the slightly more commonly used $T$ and $S$ parameters are related by $\hat T=\alpha T$ and $4s_w^2\,\hat S=\alpha S$. 

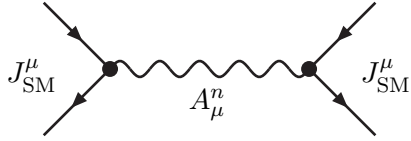
\begin{figure}[t]
\begin{center}
\begin{picture}(65,80)(35,-20)
\SetWidth{1.}
\Vertex(25,25){3}
\Vertex(100,25){3}
\ArrowLine(0,50)(25,25)
\ArrowLine(25,25)(0,0)
\Photon(25,25)(100,25){3}{5}
\ArrowLine(125,50)(100,25)
\ArrowLine(100,25)(125,0)
\put(55,10){$A_\mu^n$}
\put(-14,20){$J_{\rm SM}^\mu$}
\put(120,20){$J_{\rm SM}^\mu$}
\end{picture}
\vspace{-1cm}
\end{center}
\caption{Diagram contributing to the effective Lagrangian.}
\label{diagramas}
\end{figure}

The tree level contributions to these coefficients from integrating out the gauge KK modes of the $W$ and $Z$ bosons  can be computed from the vertex Eq.~(\ref{vertex}) according to the diagram in Fig.~\ref{diagramas}. This leads to operators consisting of products of two currents (fermionic ones as well as Higgs). The resulting effective Lagrangian contains the first operator in Eq.~(\ref{oblique}), but not the other three. 
Let us assume that the fermion fields are all localized in the UV region of the bulk. Then their coupling to the KK modes is to a large degree universal, and
 one can use the SM EOM to eliminate the fermion currents (For details see \cite{Cabrer:2011fb,Davoudiasl:2009cd})
\be
J^\mu_{\rm fermion}=-D_\nu F^{\nu\mu}-J^{\mu}_{\rm Higgs} 
\label{redef}
\ee
Now the only operators are those appearing in Eq.~(\ref{oblique}), and this basis of fields is referred to as the oblique basis. It is not hard to convince oneself that
\bea
J_{\rm Higgs}\cdot J_{\rm Higgs}&{\rm contributes\ to}& T\,,\nonumber\\
J_{\rm Higgs}\cdot J_{\rm fermion}&{\rm contributes\ to}& S,\ T\,,\nonumber\\
J_{\rm fermion}\cdot J_{\rm fermion}&{\rm contributes\ to}& S,\ T,\ Y,\ W\,.
\label{contributions}
\eea
According to our assumtions that the Higgs is near-IR and the fermions are near-UV localized, one thus finds the following parametric dependence from Eq.~(\ref{gn}).
\bea
 T&\sim&(ky_1)\,\epsilon^2\,,\nonumber\\
 S&\sim&\epsilon^2\,,\nonumber\\
W,Y&\sim&(ky_1)^{-1} \epsilon^2\,,
\label{oblique2}
\eea
where $\epsilon=m_W/m_{\rm KK}$ is the little hierarchy. One thus expects that the $T$ parameter provides the strongest contraints on $\epsilon$. Indeed, for a strictly IR localized Higgs ($a=\infty$), one finds \cite{ref} $m_{\rm KK}\gtrsim 12$ TeV. Delocalizing the Higgs into the bulk reduces the coupling to the KK modes slightly. 
Recall that the lowest value of $a$ still consistent with the RS solution to the hierarchy problem is $a=2$. This reduces $T$ by a factor of 3 and hence one still needs $m_{\rm KK}\gtrsim 7$ TeV.~\cite{Cabrer:2011fb} These bounds refer to a light Higgs boson of $m_h\sim 115$ GeV.
Recently it has been pointed out that with a heavy Higgs boson these bounds can be further reduced, due to a partial cancellation of the radiative Higgs contribution to $T$ with the KK tree level contribution. For instance, for $m_h=450$ GeV, the bound for a localized~\cite{Casagrande:2008hr} (bulk~\cite{Cabrer:2011vu}) Higgs field is $m_{\rm KK}\gtrsim 8$ TeV (4.6 TeV)

In any event, multi-TeV KK masses generate a serious little hierarchy problem and render the theory less natural. The rather large contribution to $T$ is a consequence of the fact that in the simplest model with just the SM gauge group in the bulk, the gauge-KK sector breaks the custodial symmetry of the SM at tree level. Recall that in the SM, in the limit of vanishing hypercharge  and Yukawa couplings, the Higgs sector enjoys a global "custodial" $SU(2)_R$ symmetry~\footnote{Sometimes the term "custodial symmetry" is reserved for the surviving global symmetry $SU(2)_V\subset SU(2)_L\times SU(2)_R$ in the broken phase. Since we are writing effective dimension-six operators at the scale $m_{\rm KK}\gg m_W$, it makes more sense to work in the symmetric phase and we will hence refer to $SU(2)_R$ as the custodial symmetry.} which forbids the first operator in Eq.~(\ref{oblique}). Since hypercharge does not commute with $SU(2)_R$, the $T$ parameter is generated at one-loop in the SM and its extensions. However, in the RS model there are also KK modes of the hypercharge gauge boson, and they are the culprits that generate the $T$ parameter when integrated out at tree level.~\footnote{In fact, it is a simple exercise to verify that only the hypercharge currents but not the $SU(2)_L$ currents contribute to $T$ in Eq.~(\ref{contributions}).} There are at least two ways to remedy this situation. The first~\cite{Agashe:2003zs} is to enlarge the 5D bulk gauge symmetry according to $U(1)_Y\hookrightarrow SU(2)_R$.~\footnote{An extra $U(1)$ symmetry is needed to correctly assign hypercharge to the fermions if the latter propagate in the bulk.} This embeds the hypercharge KK modes into full $SU(2)_R$ multiplets and kills any tree level contributions to $T$.  In order to achieve just the SM in the zero mode sector, one projects out the extra gauge bosons by giving them Dirichlet boundary conditions at the UV brane. Let us briefly comment on the CFT-dual interpretation of this idea. According to table \ref{tab} we have an exact global symmetry $SU(2)_L\times SU(2)_R$, of which the SM subgroup is gauged. The Higgs boson is now chosen to be a bulk or brane field transforming in the bifundamental and spontaneously breaks the global symmetry to the diagonal $SU(2)_V$. The bounds are dominated by the $S$ parameter and one can roughly achieve $m_{\rm KK}\gtrsim 3$ TeV.
The second possibility is to slightly decouple the dangerous hypercharge KK modes from the Higgs field.~\cite{Cabrer:2011fb} We will describe such a model in the next section. For yet another idea, see~\cite{Davoudiasl:2002ua}.

So far we have assumed that all fermions are near UV localized. This is a good approximation for all the light fermions of the standard model, but not so for the heavy quarks of the third generation. In order to generate a top Yukawa coupling of order unity one needs to have the left handed quark doublet to be near IR localized in order to maximize the overlap with the Higgs wave function. One thus expects a volume enhanced correction $\delta g_{Zb\bar b}\sim (k y_1)\,\epsilon^2\,$ which contributes in particular to the partial width of the $Z$ boson. For some ideas how to deal with this see \cite{Agashe:2006at}.

\section{Warped Electroweak breaking without custodial symmetry.}

We would like to suppress the KK contribution to the $T$ parameter by decreasing their coupling to the Higgs boson. Let us therefore consider the effective Higgs Lagrangian
\be
\mathcal L_{Higgs}
  =-|D_\mu H|^2+\mu^2-\lambda |H|^4+g_n J^\mu_{\rm Higgs}\cdot A_\mu^n
\ee
We will generalize the metric according to the replacement
\be
e^{ky}\to e^{A(y)}\,,
\ee
and impose $A(0)=0$, $A(y_1)=35$ in order to generate the Planck-Weak hierarchy. We will give an explicit form for $A(y)$ below, but what we have in mind is a deformation of the form
\be
A(y)=ky+({\rm  correction\ near\ }y=y_1)\,,
\ee
such that the space is asymptotically AdS near the UV brane.
By carefully integrating over the 5D Lagrangian with the Higgs zero mode profile $h(y)$ we obtain the parametric behaviour
\bea
\mu^2&\sim& Z^{-1}\rho^2\,,\nonumber\\
\lambda&\sim& Z^{-2}\,,\nonumber\\
g_n&\sim& Z^{-1}\sqrt{ky_1}\,.
\eea
where we have suppressed only $\mathcal O(1)$ quantities and defined
\be
\rho=k e^{-ky_1}\,,\qquad Z=k\int_0^{y_1}\left(\frac{h(y)\,e^{-A(y)}}{h(y_1)\,e^{-A(y_1)}}\right)^2\,.
\label{Z}
\ee
This integral arises as a wave function renormalization in the effective theory when integrating over the 5D Higgs kinetic-term. The reason we have kept it here explicitly is that under certain circumstances it can become large and suppress the coupling $g^{\rm}_n$. 
Including the $Z$ factor in the coupling $g_n$, one can see that Eq.~(\ref{oblique2}) is replaced by
\bea
 T&\sim&(ky_1)\,Z^{-2}\,\epsilon^2\,,\nonumber\\
 S&\sim&Z^{-1}\,\epsilon^2\,,\nonumber\\
W,Y&\sim&(ky_1)^{-1} \epsilon^2\,.
\label{oblique3}
\eea
In fact, it is easy to see that in pure RS $Z$ it is given by (for $a>1$)
\be
Z=\frac{1}{2(a-1)}\,.
\ee
As expected, one can gradually decouple the Higgs from the KK modes by decreasing $a$, but one is at the same time required to keep $a>2$ in order to maintain the RS solution to the hierarchy problem. One thus finds $Z_{\rm RS}<\frac{1}{2}$ which is not particularly large.

Next we will consider an explicit metric deformation of AdS in the IR region. It contains a stabilizing field $\phi$ which leads to the metric~\cite{Cabrer:2009we}
\begin{equation}
A(y)=ky-\frac{1}{\nu^2}\log\left(1-y/y_s\right) \,,
\label{metrica}
\end{equation}
where $\nu$ is a real parameter. The metric has a \allowbreak spurious singularity located at $y_s=y_1+\Delta$, outside the physical interval. It was originally studied in the absence of an IR brane in which case the singularity is a physical one and certain constraints apply to the parameter $\nu$ in order to have a sensible theory and generate a mass gap.~\cite{Cabrer:2009we} Here the singularity is shielded and $\nu$ remains arbitrary. In the limit of $\nu\to\infty$ one recovers the pure AdS metric.
In order to solve the hierarchy problem we fix $A_1\nobreak=\nobreak A(y_1)\sim\nobreak35$, which determines implicitely $ky_1<A_1$ in terms of the other parameters.

One can choose a suitable ($\phi$ dependent) bulk mass to achieve again $h(y)\sim e^{ak(y-y_1)}$.
The analogue of the condition $a>2$ in RS becomes $a>a_0$ where $a_0=2A_1/(ky_1)>2$.
The fact that $a_0>2$ can be understood in the 4D dual interpretation. 
The dimension of the Higgs condensate corresponding to the solution $h(y)\sim e^{aky}$ depends on $y$. Since the renormalization group (RG) scale is given by the warp factor we have
\be
\dim(\mathcal O_H)=\frac{h'}{h\,A'}=\frac{a}{1+\frac{1}{k(y_s-y)\nu^2}}\,.
\ee
Starting in the UV with $\dim(\mathcal O_H)>2$ the mass operator $|\mathcal O_H|^2$ has dimension~\footnote{We use the fact that in the large $N_c$ limit operator products become trivial.} $\dim(|\mathcal O_H|^2)=2\dim(|\mathcal O_H|)>4$  which, being an irrelevant operator, will become more and more suppressed along the RG flow. However following the RG flow further the theory departs from the conformal fixed point, $\dim(\mathcal O_H)$ decreases and there will be a critical RG scale $\mu_c$ at which $\dim(\mathcal O_H)<2$. As a consequence $|\mathcal O_H|^2$ will become a relevant operator and will start increasing again. As long as this happens far enough in the IR there is no concern as, at the scale $\mu_c$, the mass term is really small and there is simply not enough RG time for it to become large enough before EWSB occurs. One thus has to choose $a_0\approx\dim \mathcal O_H|_{UV}$ to be sufficiently greater than 2 such that the coefficient of $|\mathcal O_H|^2$ stays small all the way to the EW scale. The hierarchy problem is thus solved despite the fact that at the EW scale the Higgs condensate has a small dimension.

\begin{figure}[t]
\begin{center}
\begin{tabular}{ll}
\includegraphics[width=0.48\textwidth]{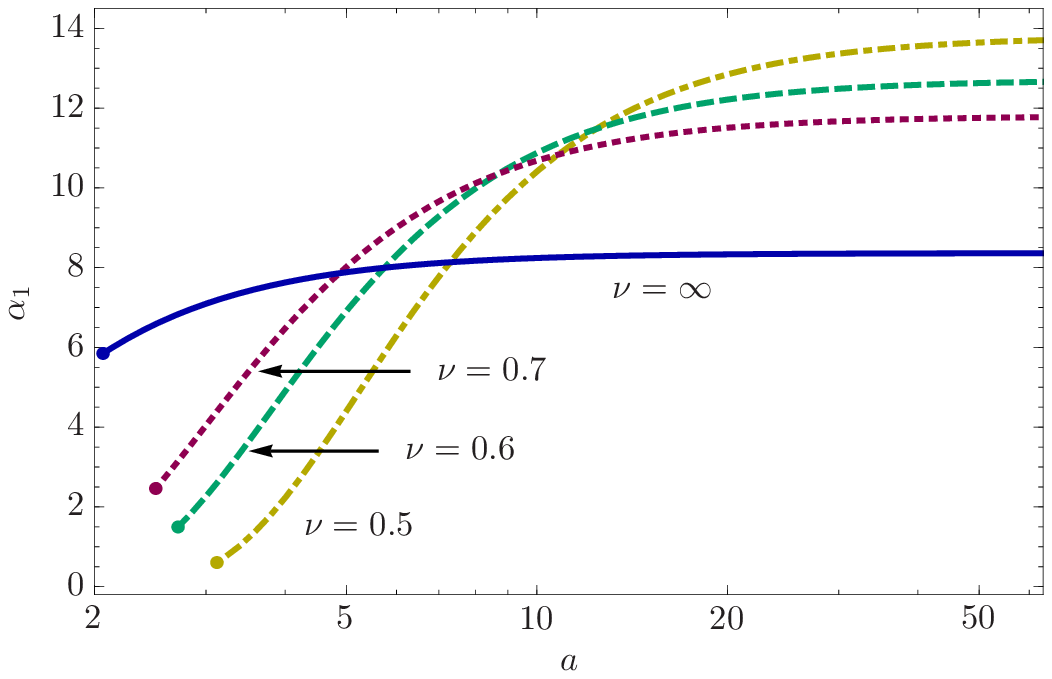}&
\input{ellipse-psfrag.tex}
\includegraphics[width=0.49\textwidth]{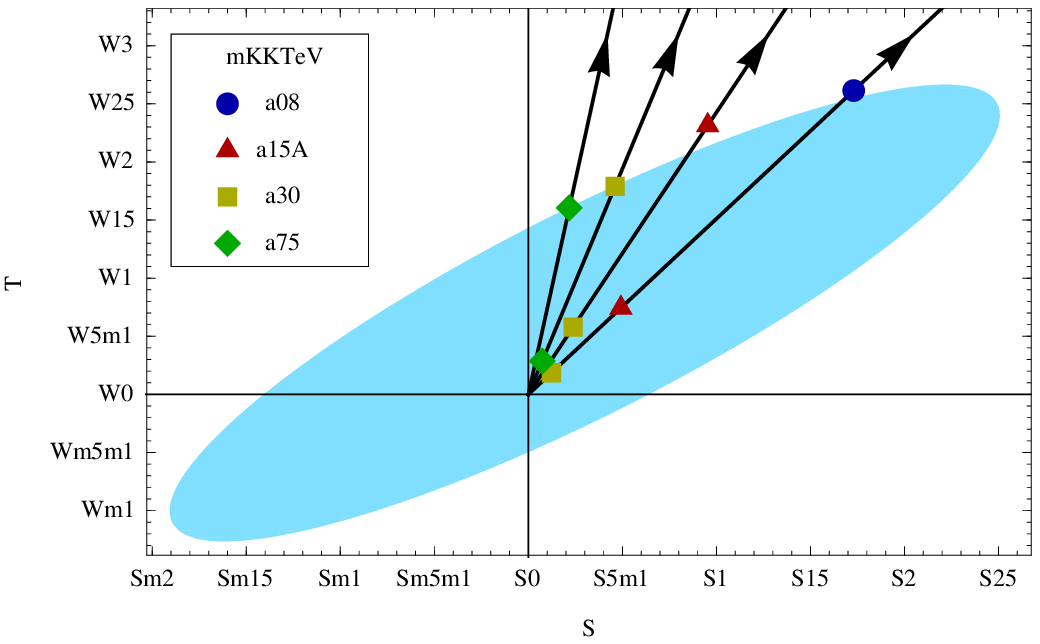}
\end{tabular}
\end{center}
\caption{
\label{fig:coupling}  
 Left panel: Plot of the coupling $\alpha_1=g_1/g$ as a function of $a$ for $\nu=0.5,0.6,0.7$ and $\infty$ (RS).  Lines end at $a=a_0$. Right panel:  95\% CL regions in the $(S,T)$ plane for different deformations. The arrows indicate decreasing KK mass.}
\end{figure}

It is clear by looking at Eq.~(\ref{Z}) that in general $Z$ will increase for decreasing $a$.
In Fig.~\ref{fig:coupling} we plot the exact coupling of the first KK mode for different strengths of the deformation as a function of the parameter $a$, computed by numerically evaluating the wave function overlap in Eq.~(\ref{gn}). One sees that, even though one has to stop at $a_0>2$, the coupling to the KK modes can be significantly reduced wrt.~RS (the line with $\nu=\infty$).
This is indeed due to the increasing $Z$ factors. 
Accordingly, the bounds are largely affected by the deformation.~\footnote{It has been noted previously  that, in models with custodial symmetry, the bounds from the $S$ parameter can be reduced with an IR-modified metric.~\cite{Fabbrichesi:2008ga}}
This is shown in the right panel of Fig.~\ref{fig:coupling}, where we have parametrized the deformation by the ratio of curvature radii in the UV ($L_0=k^{-1}$) and IR ($L_1$).~\cite{Cabrer:2011fb}
The rightmost ray corresponds to roughly the parameters $k\Delta=1$, $\nu=0.5$, i.e.~the endpoint of the yellow dashed-dotted line in the left panel of the figure. One concludes that allowing for moderate deformations from the AdS metric in the IR, the bounds on the $KK$ scale can be reduced to $\sim 1$ TeV, opening up the possibility of discovering KK resonances of the SM gauge bosons at the LHC.

A last word is in order regarding the other precision observables, in particular the $W$ and $Y$ parameters constrained by LEP-2 data, as well as the $Zb\bar b$ vertex. The former two multiply operators in Eq.~(\ref{oblique}) that do not involve the Higgs and hence will be unaffected by the reduced coupling. We have verified a posteriori that they are indeed inside their experimental errors at least in the parameter range displayed in Fig.~\ref{fig:coupling}. In contrast the $Zb\bar b$ vertex is affected by electroweak breaking and it is expected to scale as
\be
\delta g_{Zb\bar b}\sim (k y_1)\, Z^{-1}\epsilon^2\,.
\ee
Clearly one can expect some suppression due to the reduced Higgs coupling to the KK modes.
However, it is more model dependent as the precise profile of the $b$ quark has to be specified and one has to pay attention to the constraint that a large enough top Yukawa must be generated. 
Very recently it has been reported that in the described model one can achieve $m_{\rm KK}\sim 1-3$ TeV in some scenarios of fermion localization.~\cite{Carmona:2011ib}  
 
\section*{Acknowledgments}
I would like to thank the organizers of the Moriond Electroweak meeting 2011 for organizing a great conference and for inviting me to give this talk.
This work is supported by the ERC Advanced Grant 226371, the ITN programme PITN- GA-2009-237920 and the IFCPAR CEFIPRA programme 4104-2.

\end{document}